\documentclass[
reprint,
superscriptaddress,
preprintnumbers,
nofootinbib,
amsmath,amssymb,
aps,
prresearch,
floatfix,
longbibliography
]{revtex4-2}

\usepackage{graphicx}
\usepackage{dcolumn} 
\usepackage{bm}      
\usepackage{amsmath,amssymb,amsfonts,amsthm}
\usepackage{xcolor}
\usepackage[separate-uncertainty=true]{siunitx}
\usepackage{physics}
\usepackage{booktabs}
\usepackage[colorlinks=true, allcolors=blue]{hyperref}
\usepackage[nameinlink]{cleveref}

\newcommand{\secprl}[1]{\par\medskip\noindent\textbf{\emph{#1.---}}}

\AtBeginDocument{\RenewCommandCopy\qty\SI}

\Crefname{section}{Sec.}{Sec.}
\Crefname{figure}{Fig.}{Figs.}
\Crefname{table}{Tab.}{Tabs.}

\newcommand{\reply}[1]{\textcolor{black}{ #1}}

\begin{document}

\preprint{Liang, et al., \href{https://doi.org/10.1103/ykfh-cdzk}{Physical Review Research (2026)}}

\title{Estimating Orbital Parameters of Direct Imaging Exoplanet Using Neural Network}

\author{Bo Liang}
\thanks{These authors contributed equally to this work.  \\ \href{mailto:hanlin@stu.pku.edu.cn}{hanlin@stu.pku.edu.cn}}
\affiliation{Center for Gravitational Wave Experiment, National Microgravity Laboratory, Institute of Mechanics, Chinese Academy of Sciences, Beijing 100190, China}
\affiliation{Taiji Laboratory for Gravitational Wave Universe (Beijing/Hangzhou), University of Chinese Academy of Sciences (UCAS), Beijing 100049, China}

\author{Chang Liu}
\thanks{These authors contributed equally to this work.  \\ \href{mailto:hanlin@stu.pku.edu.cn}{hanlin@stu.pku.edu.cn}}
\affiliation{National Space Science Center, Chinese Academy of Sciences, Beijing 100190, China}

\author{Hanlin Song}
\thanks{These authors contributed equally to this work.  \\ \href{mailto:hanlin@stu.pku.edu.cn}{hanlin@stu.pku.edu.cn}}
\affiliation{Center for Gravitational Wave Experiment, National Microgravity Laboratory, Institute of Mechanics, Chinese Academy of Sciences, Beijing 100190, China}
\affiliation{School of Physics, Peking University, Beijing 100871, China}

\author{Tianyu Zhao}
\email{zhaotianyu@imech.ac.cn}
\affiliation{Center for Gravitational Wave Experiment, National Microgravity Laboratory, Institute of Mechanics, Chinese Academy of Sciences, Beijing 100190, China}

\author{Yuxiang Xu}
\affiliation{Center for Gravitational Wave Experiment, National Microgravity Laboratory, Institute of Mechanics, Chinese Academy of Sciences, Beijing 100190, China}
\affiliation{Shanghai Institute of Optics and Fine Mechanics, Chinese Academy of Sciences, Shanghai 201800, China}

\author{Zihao Xiao}
\affiliation{Center for Gravitational Wave Experiment, National Microgravity Laboratory, Institute of Mechanics, Chinese Academy of Sciences, Beijing 100190, China}
\affiliation{National Space Science Center, Chinese Academy of Sciences, Beijing 100190, China}

\author{Manjia Liang}
\affiliation{Center for Gravitational Wave Experiment, National Microgravity Laboratory, Institute of Mechanics, Chinese Academy of Sciences, Beijing 100190, China}

\author{Minghui Du}
\affiliation{Center for Gravitational Wave Experiment, National Microgravity Laboratory, Institute of Mechanics, Chinese Academy of Sciences, Beijing 100190, China}

\author{Wei-Liang Qian}
\affiliation{Escola de Engenharia de Lorena, Universidade de S\~ao Paulo, Lorena, SP 12602-810, Brazil}

\author{Li-e Qiang}
\affiliation{Center for Gravitational Wave Experiment, National Microgravity Laboratory, Institute of Mechanics, Chinese Academy of Sciences, Beijing 100190, China}

\author{Mingming Sun}
\affiliation{AGI Lab, Beijing Institute of Mathematical Sciences and Applications, Beijing, China}

\author{Peng Xu}
\email{xupeng@imech.ac.cn}
\affiliation{Center for Gravitational Wave Experiment, National Microgravity Laboratory, Institute of Mechanics, Chinese Academy of Sciences, Beijing 100190, China}
\affiliation{Key Laboratory of Gravitational Wave Precision Measurement of Zhejiang Province, Hangzhou Institute for Advanced Study, UCAS, Hangzhou 310024, China}
\affiliation{Lanzhou Center of Theoretical Physics, Lanzhou University, Lanzhou 730000, China}

\author{Ziren Luo}
\affiliation{Center for Gravitational Wave Experiment, National Microgravity Laboratory, Institute of Mechanics, Chinese Academy of Sciences, Beijing 100190, China}
\affiliation{Key Laboratory of Gravitational Wave Precision Measurement of Zhejiang Province, Hangzhou Institute for Advanced Study, UCAS, Hangzhou 310024, China}

\date{\today}

\begin{abstract}
We propose a flow-matching Markov chain Monte Carlo (FM-MCMC) algorithm for estimating the orbital parameters of exoplanetary systems, especially for those only one exoplanet is involved. Compared to traditional methods that rely on random sampling within the Bayesian framework, our approach first leverages flow matching posterior estimation (FMPE) to efficiently constrain the prior range of physical parameters, and then employs MCMC to accurately infer the posterior distribution. For example, in the orbital parameter inference of $\beta$ Pictoris b, our model achieved a substantial speed-up while maintaining comparable accuracy—running 77.8 times faster than Parallel Tempered MCMC (PTMCMC) and 365.4 times faster than nested sampling. Moreover, our FM-MCMC method also attained the highest average log-likelihood among all approaches, demonstrating its superior sampling efficiency and accuracy. This highlights the scalability and efficiency of our approach, making it well-suited for processing the massive datasets expected from future exoplanet surveys. Beyond astrophysics, our methodology establishes a versatile paradigm for synergizing deep generative models with traditional sampling, which can be adopted to tackle complex inference problems in other fields, such as cosmology, biomedical imaging, and particle physics.
\end{abstract}

\maketitle

\secprl{Introduction}
Since the first confirmed detection of an exoplanetary system in 1992 \cite{Wolszczan1992} and an exoplanet in 1995 \cite{1995Natur.378..355M}, more than 4,615 exoplanetary systems and 5,989 exoplanets have been confirmed as of September 2025 \cite{nasa_exoplanet_archive_2025}. These systems exhibit a wide range of characteristics, including variations in the number and composition of exoplanets, the dynamics of their orbital motions, and the types of their host stars. Notably, the majority of these systems contain only one confirmed planet, a pattern that is partly attributable to observational biases. For a recent review, see, e.g., Ref. \cite{howe2025architecture}. High-precision exoplanet detectors are essential for investigating the mechanisms underlying these characteristics, such as testing theories of planet formation and migration, and understanding the overall evolution of planetary systems \cite{Brandt_2021, Blunt_2020}. Currently, several observatories remain in operation and are continuously expanding our exoplanet census. Operating projects like Transiting Exoplanet Survey Satellite (TESS)~\cite{winn2024transitingexoplanetsurveysatellite} and CHaracterising ExOPlanets Satellite (CHEOPS)~\cite{fortier2024cheopsinflightperformancecomprehensive} conduct high-precision photometric surveys of nearby bright stars, enabling the detection and characterization of transiting exoplanets. In parallel, Gaia~\cite{PANCINO20201} ultra-precise astrometric data that reveals subtle stellar motions caused by orbiting planets. Additionally, instruments like Spectro-Polarimetric High-contrast Exoplanet REsearch (SPHERE)~\cite{beuzit2019sphere} on the Very Large Telescope (VLT)  directly image exoplanets, offering valuable constraints on their orbital architectures and atmospheric properties. Upcoming missions such as PLAnetary Transits and Oscillations of stars (PLATO)~\cite{rauer2024platomission}, Atmospheric Remote-sensing Infrared Exoplanet Large-survey (ARIEL)~\cite{changeat2023esaarieldatachallengeneurips,Zingales2017TheAM}, SPHERE+ \cite{2020arXiv200305714B}, and Nancy Grace Roman Space Telescope~\cite{bailey2023nancygraceromanspace} are expected to significantly advance our understanding of exoplanet populations. By integrating large-scale photometric monitoring, atmospheric spectroscopy, and precise astrometric measurements, these missions are poised to deliver unprecedented insights into the formation, evolution, and diversity of planetary systems. In recent years, a variety of exoplanet discovery techniques have been developed and extensively employed \cite{maire2023workshop}, including transit photometry, radial velocity (RV), microlensing, and direct imaging. These efforts have led to the detection of a large number of exoplanetary systems, significantly advancing our understanding of the statistical properties of planetary populations and providing critical constraints on models of planet formation and migration. In this work, we focus on exoplanetary systems detected using direct imaging method. Although the number of such detections is smaller than that obtained with the other three techniques \cite{2014prpl.conf..715F, 2018haex.bookE....D}, direct imaging provides unique advantages for determining the demographics of exoplanets around nearby young stars, characterizing their atmospheric properties, and exploring their circumstellar disks \cite{2020arXiv200305714B}. Through synergy with other facilities, such as Gaia, GRAVITY \cite{abuter2017first}, and James Webb Space Telescope (JWST)~\cite{lightsey2012james}, more exoplanets will be directly imaged. This growing catalog of directly imaged exoplanets will yield essential observational data, which is vital for testing and validating theoretical models of planet formation and evolution, thereby elucidating the fundamental physical processes that determine planetary system architectures.

With the growing accumulation of observational data, a key challenge lies in the rapid and precise estimation of the physical parameters of exoplanetary systems, which are crucial for predicting exoplanet locations, calculating transit probabilities, and assessing the climates and habitability of Earth-like worlds in future missions. In the case of directly imaged exoplanets, several approaches have been proposed for estimating the physical parameters \cite{Brandt_2021}, such as linearized least square adjustment \cite{forveille1999accuratemasseslowmass}, the Orbits for the Impatient (OFTI) algorithm \cite{2017AJ....153..229B} and the Parallel Tempered Markov Chain Monte Carlo (PTMCMC) algorithm \cite{2013PASP..125..306F, 2016MNRAS.455.1919V}. Tools such as \texttt{orbitize!} implement Nested Sampling~\cite{10.1093/mnras/staa278}, and have been demonstrated in studies that combine astrometric and radial velocity data to constrain exoplanet orbits \citep{Brandt_2021, Blunt_2020}. However, these traditional methods~\cite{forveille1999accuratemasseslowmass,2017AJ....153..229B,2013PASP..125..306F, 2016MNRAS.455.1919V} suffer from limitations: Due to high parameter dimensionality and the potential multimodality of posterior distributions, convergence can be prohibitively slow without an optimal proposal distribution or favorable random initialization. In contrast, while machine learning based methods~\cite{Sun_2022, ruth_exoplanet_2024, chen2024deepttv} offer a significant advantage in speed for rapid orbital parameter inference, the posterior distributions they generate typically exhibit larger errors than traditional MCMC methods and often fail to capture multimodal features accurately. Consequently, achieving rapid inference of exoplanet orbital parameters while maintaining high precision remains a pressing challenge.

In this work, we present a  approach, \emph{flow-matching enhanced MCMC} (FM-MCMC), which leverages continuous normalizing flows (CNFs)~\cite{chen2021eventfn} trained via a flow-matching objective~\cite{liu2023flow,dax2023flowmatchingscalablesimulationbased,lipman2023flow} to accelerate MCMC sampling. Our method achieves posterior estimates that closely approximate those obtained by PTMCMC and nested sampling, while outperforming neural posterior estimation~\cite{NIPS2016_6aca9700, lueckmann2017flexiblestatisticalinferencemechanistic} in precision. FM‑MCMC combines the robustness and interpretability of traditional MCMC sampling with the computational speed of normalizing flow, yielding higher posterior accuracy than standalone ---\reply{simulation-based inference} methods~\cite{doi:10.1073/pnas.2109420119, Wong:2022xvh, Gabrie:2021tlu}.

\secprl{Methodology}
In this section, we introduce the FM-MCMC algorithm to accelerate posterior sampling. Based on the CNFs and flow matching algorithm, we first train a neural network for initial inference. Then, along with the PTMCMC, we obtain the accurate posterior distributions. Unlike traditional MCMC methods—which often suffer from slow convergence in high-dimensional and multi-modal parameter spaces due to inefficient proposal mechanisms and sensitivity to initial conditions—FM-MCMC leverages a learned flow-based proposal that drastically reduces the number of likelihood evaluations and shortens burn-in phases.

\begin{figure}[t]
  \centering
  \includegraphics[width=0.48\textwidth]{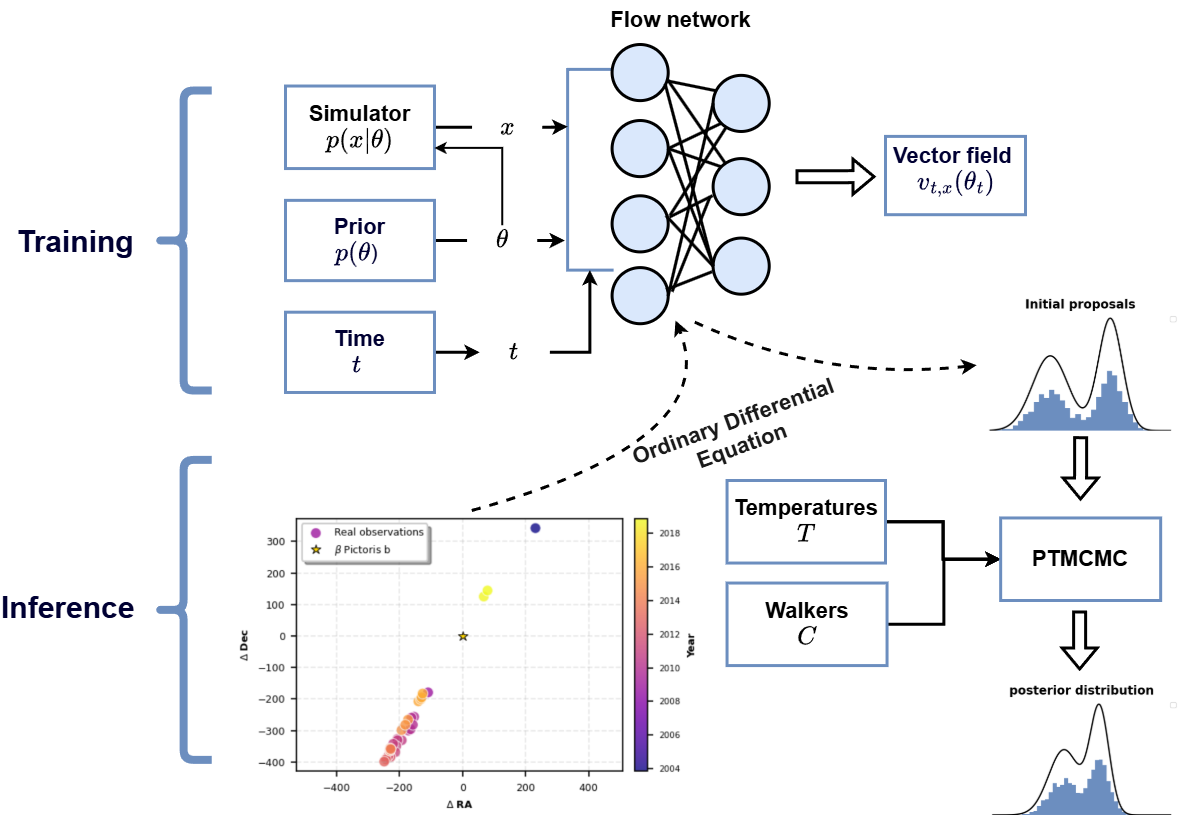}
  \caption{\textbf{Flow-matching MCMC framework.} In the Training section, there is first a prior distribution $p(\theta)$, from which the parameter $\theta$ is extracted, and the data is generated by a model $x$. These generated data are fed into the training process along with the parameters. The inference section is based on the trained model, input the real observation data, and use the trained model to infer the initial proposals. The initial proposal is then provided to PTMCMC for likelihood calculations.}
  \label{fig:model}
\end{figure}

\emph{Continuous normalizing flows.}---Generative models are designed to learn invertible transformations that map high-dimensional data distributions to simpler reference distributions. Notably, Normalizing Flows achieve exact density estimation by applying the chain rule to compositions of invertible functions. However, their discrete layer-wise architecture limits their capacity to model highly complex data distributions. To overcome this limitation, CNFs were introduced. CNFs construct a time-dependent probability density path \( p_t(\theta) \) governed by a neural network-parameterized vector field \( v_t(x; \theta) \). The evolution of samples in CNFs is described by an ordinary differential equation (ODE):
\begin{equation}
\frac{d}{dt}\phi_{t,x}(\theta)=v_{t,x}(\phi_{t,x}(\theta)),\quad v_{t=0,x}(\theta)=\theta_{0} ,\quad v_{t=1,x}(\theta)=\theta_{1}.
\end{equation}

Traditional methods optimize the model via maximum likelihood estimation by minimizing the negative log-likelihood. However, backpropagation through an ODE solver incurs a computational cost of \( \mathcal{O}(d^3) \), where \( d \) denotes dimensionality, hindering scalability to high-dimensional data. FM offers a solution by introducing a more efficient optimization objective. In contrast to conventional CNFs, FM decouples the probability density path from the underlying dynamics, substantially reducing computational overhead. It explicitly constructs a target probability path \( p_t(\theta) \) and its associated velocity field \( u_t(\theta) \), and trains a parametric velocity field \( v_{t,x}(\theta)\) to approximate \( u_t(\theta) \). Dax et al. applied FM to simulation-based inference, proposing a highly efficient training objective that alleviates the computational demands of traditional methods. The method begins by defining a Gaussian probability path conditioned on a target parameter \( \theta_1 \):
\begin{equation}
    p_t(\theta \mid \theta_1) = \mathcal{N}\left( \theta| \theta_1 t, (1 - t)^2 I_d \right),
\end{equation}
which ensures the distribution starts as a standard normal \( \mathcal{N}(0, I_d) \) at \( t = 0 \) and converges to a Gaussian centered at \( \theta_1 \) when \( t = 1 \). The corresponding conditional velocity field that generates this path is derived as:
\begin{equation}
    u_t(\theta \mid \theta_1) = \theta_1 - \theta,
\end{equation}
enabling linear transport of samples from the base to the target distribution over time. To approximate the posterior, the framework marginalizes over target parameters \( \theta_1 \sim p(\theta) \) and conditions on observed data \( x \sim p(x \mid \theta_1) \), yielding an aggregate probability path \( p_t(\theta) \) and a composite velocity field \( u_t(\theta) \) such that \( p_1(\theta) = q(\theta \mid x) \). The training objective minimizes the mean-squared error between the learned and true velocity fields:
\begin{equation}
\begin{split}
    \mathcal{L}_{\text{FMPE}} = \, & \mathbb{E}_{t \sim \mathcal{U}(0,1)} \, \mathbb{E}_{\theta_1 \sim p(\theta)} \, \mathbb{E}_{x \sim p(x \mid \theta_1)} \\
    & \mathbb{E}_{\theta_t \sim p_t(\theta_t \mid \theta_1)} \left\| v_{t,x}(\theta_t) - u_t(\theta_t \mid \theta_1) \right\|^2.
\end{split}
\end{equation}
This regression-based approach avoids expensive backpropagation through ODE solvers, greatly improving scalability in high-dimensional settings. Once trained, posterior samples are generated by solving the ODE numerically, effectively transforming base distribution samples into draws from the approximate posterior.

\emph{FM-MCMC Sampling.}---The FM-MCMC algorithm first employs CNFs trained through flow matching to enable rapid obtaining the initial proposals to efficiently generate initial proposals tailored to specific scientific problems. These proposals are then refined using PTMCMC to obtain accurate posterior distributions. For example, when applying the FM-MCMC algorithm to estimate the orbital parameters of imaged exoplanet, i.e. $\beta$ Pictoris b, a two-step inference is performed. Step 1: Obtain the initial proposal for the orbital parameters of $\beta$ Pictoris b using a pretrained neural network. Step 2: Obtain the accurate posterior distribution with the injected initial proposal using the PTMCMC algorithm. The CNFs sample production dynamically adapts to PTMCMC's chain configurations and temperature ladder parameters. This integration delivers two key enhancements: (1) Using CNFs posterior samples as the initialization proposal distribution, and (2) Burn-in phases are reduced to 1\% of the original requirement. Final posterior distributions are derived from thermally equilibrated PTMCMC chains.
\begin{figure}[htb]
  \centering
  \includegraphics[width=0.4\textwidth]{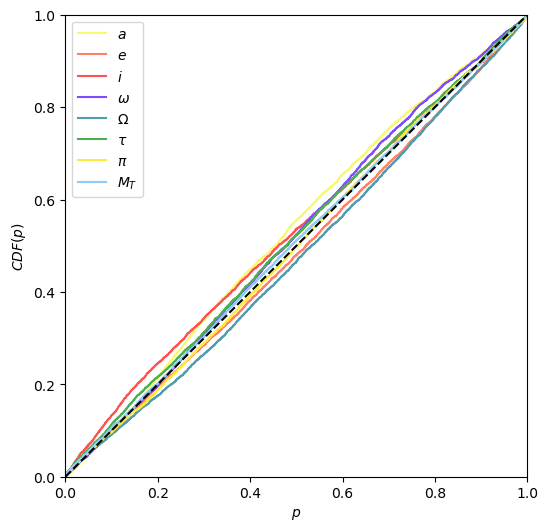}
  \caption{\textbf{P-P plot validating the unbiasedness of the CNF model.} The near-perfect alignment of all CDF curves with the theoretical diagonal indicates statistical consistency between the CNF posterior estimates and the true distribution.}
  \label{fig:pp}
\end{figure}

\emph{Data and Model.}---Since our ultimate goal is to estimate the orbital parameters of $\beta$ Pictoris b with the trained neural network, it is essential to construct a training dataset consisting of simulated observations that closely resemble systems analogous to $\beta$ Pictoris b. In this work, we generate a dataset comprising 16 million simulated orbital configurations. The relative astrometry data of exoplanets are often expressed as the star-planet separation and position angle, or equivalently as the relative right ascension ($\Delta \text{RA}$) and relative declination ($\Delta \text{Dec}$) \cite{Blunt_2020, 2020AJ....159...71N}. These observed orbital data can be parameterized with eight quantities: the six standard Keplerian elements plus parallax $\pi$ and total system mass $M_{\text{tot}}$. The specific priors and noise models used for training are detailed in the Supplemental Material.

\begin{figure}[t]
  \centering
  \includegraphics[width=0.5\textwidth]{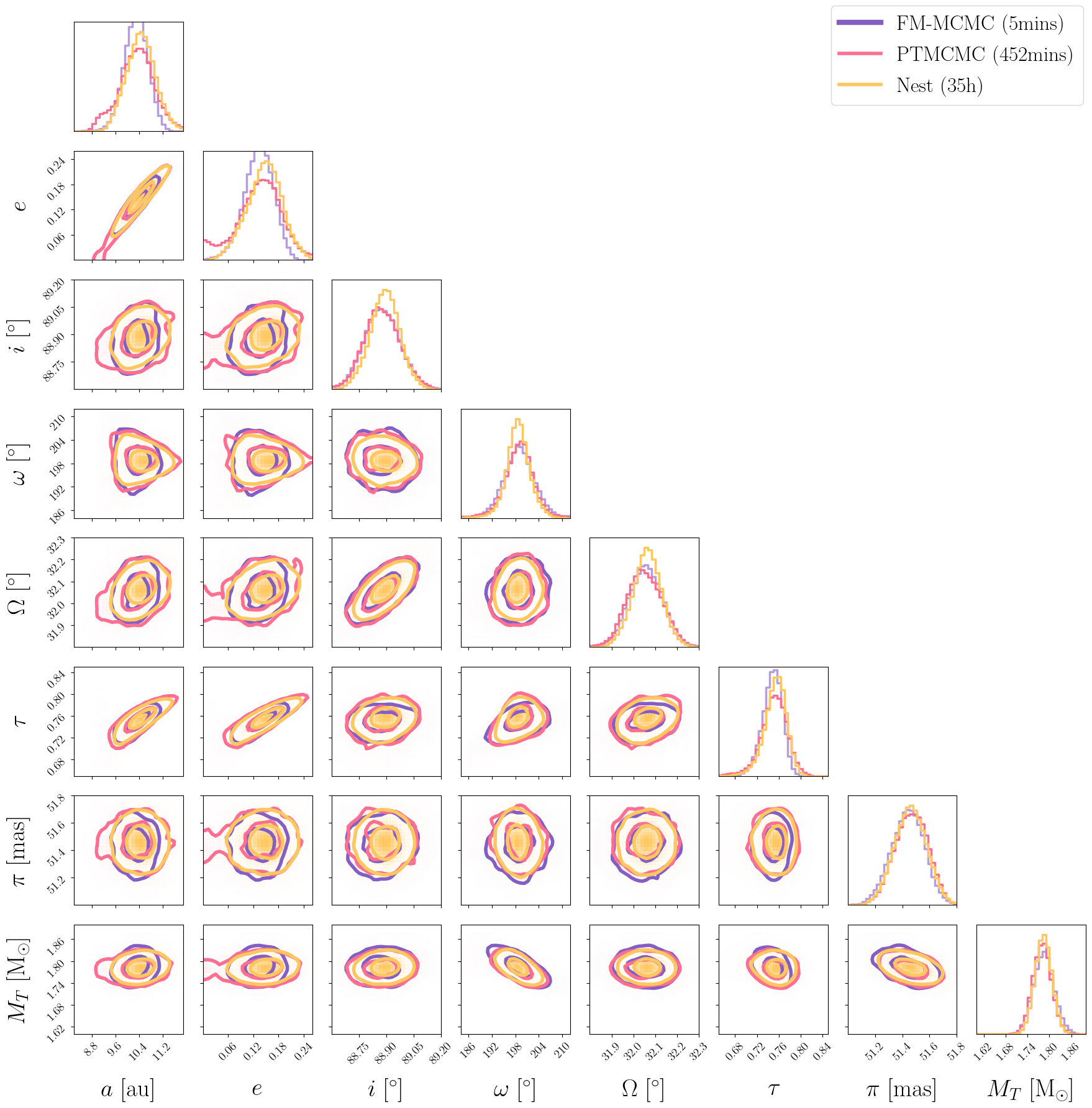}
  \caption{\textbf{FM‑MCMC vs PTMCMC vs Nested sampling posterior comparison for $\beta$-Pictoris b orbital elements.} Posterior distributions of $\beta$-Pictoris b’s orbital parameters inferred by FM‑MCMC (blue), PTMCMC (orange), and nested sampling (red) show statistical consistency, with nearly identical $1\sigma$ and $2\sigma$ credible regions.}
  \label{fig:corner_mcmc}
\end{figure}

\secprl{Results}
With the flow-matching-based neural network, the orbital parameters of $\beta$ Pictoris b can be inferred through the two-step procedure described above. Our FM-MCMC results demonstrate two major improvements over previous parameter estimation methods that rely solely on either machine learning or traditional sampling algorithms.

\emph{Efficiency Improvement.}---To rigorously evaluate the effectiveness of FM-MCMC, we performed posterior estimation for $\beta$ Pictoris b using both PTMCMC and Nested sampling. The PTMCMC sampler was configured with 20 temperature tiers and 1,000 parallel chains. After burn-in, we retained the final 20,000 samples for posterior analysis. The PTMCMC computation completed in 452 minutes, while the Nested sampling algorithm required 2124 minutes to achieve convergence.
\begin{table}[b]
\centering
    \caption{Comparison of efficiency and effectiveness of different posterior sampling approaches.}\label{table:efficiency}
        \begin{tabular*}{\columnwidth}{@{\extracolsep{\fill}}llll}
        \toprule
        \textbf{Method} & \textbf{Time (s)} $\uparrow$& \textbf{Speedup} & \textbf{Log $\mathcal{L}$} (mean) $\uparrow$\\
        \hline 
        FM-MCMC (ours) & \textbf{348.9} & 1 & \textbf{-133.0} \\
        PTMCMC & 27133.9 & 77.8 & -285.4 \\
        Nested sampling & 127478.8 &  365.4 & -234.3 \\
        \bottomrule
        \end{tabular*}
\end{table}
As shown in Table~\ref{table:efficiency}, FM-MCMC demonstrates high efficiency by integrating machine learning with traditional sampling. Under identical configurations (20 temperature tiers, 1,000 parallel chains), FM-MCMC completes sampling in \textbf{348.9 seconds} ($\approx$5.8 minutes), outperforming PTMCMC (27,133.9s/7.5h) and Nested sampling (127,478.8s/35.4h) by \textbf{77.8$\boldsymbol{\times}$} and \textbf{365.4$\boldsymbol{\times}$}, respectively. Moreover, likelihood comparisons show that FM-MCMC achieves a mean negative log-likelihood of $-133.0$, which is \textbf{53.4\% better} than that of PTMCMC ($-285.4$) and \textbf{43.2\% better} than that of Nested sampling ($-234.3$). 
\reply{Critically, its \textbf{maximum likelihood} (-129.6) equals that of PTMCMC (-129.6) and deviates by only \textbf{0.08\%} from that of Nested sampling (-129.5), demonstrating that FM-MCMC achieves enhanced efficiency without compromising posterior accuracy.}
\reply{The offline training of the FMPE network was completed in about 22 minutes on an NVIDIA RTX 4090 GPU using early stopping.}

\emph{Accuracy.}---FM-MCMC demonstrates statistical robustness in parameter inference. As shown in Fig.\ref{fig:corner_mcmc}, both the 2D contours and 1D marginal distributions of all FM-MCMC parameters are consistent with results obtained from PTMCMC and Nested Sampling, confirming the fidelity of the inferred distributions. Detailed posterior estimates and comparisons with purely neural-based methods (NPE) are provided in the Supplemental Material.

\emph{Robustness Verification.}---To statistically validate the calibration of the CNFs architecture across the full parameter space, Figure~\ref{fig:pp} presents a Probability-Probability (P-P) plot demonstrating the model’s unbiasedness, where all parameter-wise cumulative distribution functions (CDFs) align closely with the theoretical diagonal. This consistency highlights the unbiased nature of the trained model, confirming its reliability and robustness across diverse parameters.

\emph{Scientific Implications.}---\reply{The capability to reduce inference time from hours to minutes enables rapid updates of orbital posteriors for dynamically active systems. For targets like $\beta$ Pictoris b, fast orbital inference is useful for rapid follow-up observations. Newly obtained astrometric data can be incorporated quickly to update the predicted separation, position angle, and observability of a planet during future observing windows. Such rapid updates will be valuable for prioritizing follow-up spectroscopy in future large surveys of directly imaged planets.} Furthermore, the rapid posterior sampling facilitates on-the-fly N-body simulations to probe disk-planet interactions~\cite{BELLEMAN2008103, Lacquement_2025} and provides informed priors to accelerate the search for inner companions~\cite{Lacour_2021}. More broadly, this efficiency helps remove a computational bottleneck in characterizing the large exoplanet datasets expected from future surveys and astrometric missions such as Roman and Gaia.

\secprl{Conclusion}
We have presented FM‑MCMC, a flow‑matching enhanced MCMC sampler for astrometric orbit fitting, and demonstrated its application to the well‑studied $\beta$ Pictoris b system. FM‑MCMC produces posterior distributions that are statistically consistent with those obtained via fully converged PTMCMC and nested sampling, while achieving a roughly 100× reduction in wall‑clock runtime. Importantly, this substantial speed-up is achieved without sacrificing the accuracy of the posterior estimates. By integrating learned transport maps into a conventional MCMC sampler, FM‑MCMC unites the statistical rigor of Bayesian inference with the high computational efficiency characteristic of modern generative models. While our study focused on $\beta$ Pictoris b using priors consistent with previous works, FM‑MCMC shows promise for application to broader prior ranges, multi‑planet systems, and entirely new scientific problems. 

\reply{
In the present implementation, the network is trained for the specific observational setup and prior ranges adopted for the $\beta$ Pictoris b analysis. Therefore, applying the current model to another directly imaged planet with different observing epochs, measurement uncertainties, or prior ranges requires a new simulated training set and retraining. This retraining step is computationally inexpensive; for the $\beta$ Pictoris b analysis, it takes only about 22 minutes, allowing the present implementation to be efficiently adapted to other data sets. We also note that this source-specific training requirement is not an intrinsic limitation of FM-MCMC. A future generalized implementation could include observing epochs, measurement uncertainties, missing-data masks, and system-level quantities in the conditioning vector, allowing a single trained model to be applied to multiple sources and heterogeneous observation sets.
}

\textit{Acknowledgement---}
\reply{We thank the anonymous referee for profound comments.} This study is supported by the National Key Research and Development Program of China (Grant No. 2021YFC2201901, Grant No. 2021YFC2203004, Grant No. 2020YFC2200100 and Grant No. 2021YFC2201903). International Partnership Program of the Chinese Academy of Sciences, Grant No. 025GJHZ2023106GC. We also gratefully acknowledge the financial support from Brazilian agencies Funda\c{c}\~ao de Amparo \`a Pesquisa do Estado de S\~ao Paulo (FAPESP), Fundação de Amparo à Pesquisa do Estado do Rio Grande do Sul (FAPERGS), Funda\c{c}\~ao de Amparo \`a Pesquisa do Estado do Rio de Janeiro (FAPERJ), Conselho Nacional de Desenvolvimento Cient\'{\i}fico e Tecnol\'ogico (CNPq), and Coordena\c{c}\~ao de Aperfei\c{c}oamento de Pessoal de N\'ivel Superior (CAPES). This work is also supported by High-performance Computing Platform of Peking University.

\reply{\textit{Data and code availability---}
The code and trained neural-network model used in this work are publicly available in the FM-MCMC GitHub repository~\cite{Liang2025FMMCMCGitHub}.}

\bibliography{ref}

\end{document}